\documentstyle[12pt,subfigure,rotating]{article}

\newcommand{\epsfig}[1]{\fbox{ -Figures not included- }}

\newcommand{\D}{\displaystyle}

\newcommand{\bildchen}[3]{%
\begin{center}
\begin{sideways}
\makebox{\hspace*{15ex} \Large $\D #2$}
\end{sideways}
\mbox{{\epsfig{figure=#1,width=13.cm,%
bbllx=1.8cm,bblly=9.2cm,bburx=20.cm,bbury=19.cm}}}
\end{center}
\begin{flushright}
   {\Large $\D #3$ \hspace*{5ex}}
\end{flushright}}

\newcommand{\be}{\begin{equation}}
\newcommand{\ee}{\end{equation}}
\newcommand{\comment}[1]{}

\newcommand{\mcal}{{\cal M}}

\newcommand{\strich}[1]{#1  \! \! \slash}

\newcommand{\gnr}{\Gamma_{\tau \rightarrow \nu \pi}}

\begin{document}

\makebox[14cm][r]{Oktober 1994}\par
\makebox[14cm][r]{hep-ph/9411316}\par
\makebox[14cm][r]{LNF-94/067(p)}\par
\vspace{.7cm}
\centerline{\Large \bf
Radiative Corrections to the Decay $\tau\to\pi\nu_\tau$}
\par
\vspace{1.cm}
\centerline{\sc Roger Decker}
\par \vspace*{0.5cm}
\centerline{Institut f\"ur Theoretische Teilchenphysik} \par
\centerline{Universit\"at Karlsruhe}\par
\centerline{D--76128 Karlsruhe, Germany}\par\par
\vspace*{1.5cm}
\centerline{\sc Markus Finkemeier}
\par \vspace*{0.5cm}
\centerline{INFN - Laboratori Nazionali di Frascati} \par
\centerline{P.O. Box 13}\par
\centerline{00044 Frascati (Roma), Italy}\par
\par
\normalsize
\vspace{2.cm}

\begin{abstract}
We have calculated the radiative corrections to the decay
$\tau\to \pi(K) \nu_\tau$,
taking to account internal
bremsstrahlung and structure dependent radiation in the radiative
decay and point meson, hadronic structure dependent and short
distance contributions in the virtual corrections.
We display the spectra of the photon energy and of the pion-photon
invariant mass in the decay $\tau\to\pi\nu_\tau\gamma$ and compare with
the PHOTOS Monte-Carlo.
Our result for the radiative correction to the ratio
$\Gamma(\tau\to\pi\nu_\tau(\gamma))/ \Gamma(\pi\to\mu\nu_\mu(\gamma))$
is
$\delta R_{\tau/\pi} = \left(0.16 \pm 0.14 \right)  \%$
and for the ratio
$\Gamma(\tau\to K\nu_\tau(\gamma))/ \Gamma(K\to\mu\nu_\mu(\gamma))$,
we obtain
$\delta R_{\tau/K } = \left(0.90 \pm 0.22 \right)
\%$.
\\
\end{abstract}
\begin{center}
Invited talk at the Third Workshop on Tau Lepton Physics
(Montreux, September 1994)
\end{center}
\begin{center}
presented by: Markus Finkemeier
\end{center}

\newpage
\section{Introduction}
%
\subsection{Lowest Order Prediction}
%
The simplest semihadronic decay mode of the tau lepton is unique in that
there is essentially no uncertainty in the theoretical prediction at the order
$O(\alpha^0)$. The decay width $\Gamma(\tau\to\pi\nu_\tau)$ can be predicted
using the precisely measured pion decay width $\Gamma(\pi\to\mu\nu_\mu)$.
The only parameter in both decays is the pion decay constant
$f_\pi$ defined by the matrix element of the weak hadronic current,
\be
  < 0| J^\mu_{\mbox{\scriptsize weak}}| \pi^+(p)>
   = i \sqrt{2} f_\pi p^\mu
\ee
In the ratio $R_{\tau/\pi}$ of the decay rates
\be
    R_{\tau/\pi} = \frac{\Gamma(\tau\to\pi\nu_\tau)}
      {\Gamma(\pi\to\mu\nu_\mu)}
\ee
the coupling $f_\pi$ cancels, and the resulting formula involves the
masses of the particles only,
\be
   R_{\tau/\pi}
   = \frac{1}{2} \frac{m_\tau^3}{m_\pi m_\mu^2}
   \frac{(1 - m_\pi^2/m_\tau^2)^2}{(1-m_\mu^2/m_\pi^2)^2} + O(\alpha)
\ee
Multiplying this ratio
by the experimental decay width $\Gamma(\pi\to\mu\nu_\mu)$,
one gets the $O(\alpha^0)$ prediction for $\Gamma(\tau\to\pi\nu_\tau)$.
%
\subsection{Radiative Corrections of the order $O(\alpha)$}
\begin{figure}
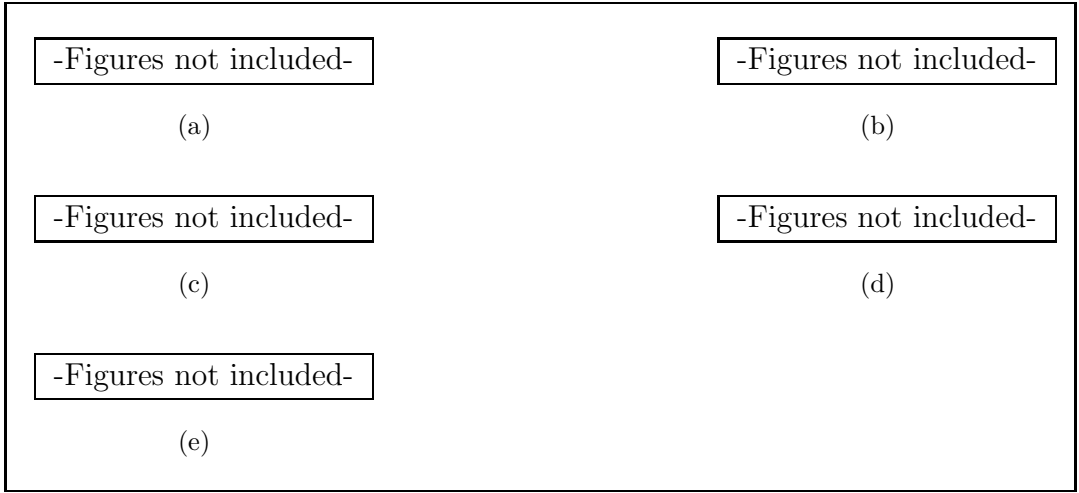

   \caption{Diagrams determining the radiative corrections to
    $\tau\to\pi\nu_\tau$}
   \label{fig2}
\begin{center}
\fbox{\parbox{14cm}{
\subfigure[]{
\mbox{   \epsfig{figure=figc1a.ps,width=4.6cm,%
bbllx=2cm,bblly=7cm,bburx=18.1cm,bbury=21.6cm}}
}
\subfigure[]{
\mbox{   \epsfig{figure=figc1b.ps,width=4.6cm,%
bbllx=2cm,bblly=7cm,bburx=18.1cm,bbury=21.6cm}}
}
\subfigure[]{
\mbox{   \epsfig{figure=figc1c.ps,width=4.6cm,%
bbllx=2cm,bblly=7cm,bburx=18.1cm,bbury=21.6cm}}
}
\subfigure[]{
\mbox{   \epsfig{figure=figb0a.ps,width=4.6cm,%
bbllx=2cm,bblly=7cm,bburx=18.1cm,bbury=21.6cm}}
}
\subfigure[]{
\mbox{   \epsfig{figure=figb0b.ps,width=4.6cm,%
bbllx=2cm,bblly=7cm,bburx=18.1cm,bbury=21.6cm}}
}
%
}}
  \end{center}
\end{figure}

%
Because of the infra-red divergences, at the order $O(\alpha)$ one has to
consider the ratio of inclusive decay rates
into final states with or without an additional photon:
\begin{eqnarray}
{   R_{\tau/\pi}    =  \frac{\Gamma(\tau\to\pi\nu_\tau(\gamma))}
      {\Gamma(\pi\to\mu\nu_\mu(\gamma))}}
    =  \frac{1}{2} \frac{m_\tau^3}{m_\pi m_\mu^2}
   \frac{(1 - m_\pi^2/m_\tau^2)^2}{(1-m_\mu^2/m_\pi^2)^2}
  \Big( 1 + \delta R_{\tau/\pi} \Big)
\end{eqnarray}
The radiative correction $\delta R_{\tau/\pi}$ arises from the general Feynman
diagrams shown in Fig.~\ref{fig2}, and the corresponding diagrams for the decay
$\pi\to\mu\nu_\mu(\gamma)$. There  are the virtual corrections, Fig.~\ref{fig2}
(a)--(c), and the corrections due to the radiative decay, (d)--(e). The
diagrams fall into three classes. The diagrams of the first class (Fig.~%
\ref{fig2} (a) and (d)), where the photon only couples to the leptonic side,
are determined by $f_\pi$ and QED, and so they can be
calculated without theoretical uncertainty. The diagrams of the third class
(Fig.~\ref{fig2} (c)), where the photon couples twice to the hadronic side,
give identical corrections for the tau and pion decay and so cancel in
$R_{\tau/\pi}$. Thus the essential problem
is to calculate the diagrams of the second
class, where the photon couples once to the hadronic side, Fig.~\ref{fig2}
(b) and (e).

What makes the calculation of these radiative corrections difficult and
interesting is the fact that due to the integration over the momentum of
the virtual photon, the physics of very different energy scales $Q$ is
involved. For very small energy scales, $Q^2 \ll m_\rho^2$, the relevant
amplitudes are fixed by low energy theorems of QCD. For intermediate scales,
$Q^2 \approx m_\rho^2$, the amplitudes are dominated by
hadronic resonances, and for large scales $Q^2 \gg m_\rho^2$, they are
dominated by the short distance behaviour of the weak interaction.

In Sec.~2 we will discuss the radiative decay, which is not only an essential
part of the total radiative correction, but also of interest in its own right.
In Sec.~3 we will then consider the virtual corrections and the total
correction to the ratio $R_{\tau/\pi}$.

\section{The Radiative Decay $\tau\to\pi\nu_\tau\gamma$}
%
The amplitude for the radiative decay $\tau\to\pi\nu_\tau\gamma$ can
be divided into the so-called internal bremsstrahlung (IB) and the
structure dependent radiation (SD) \cite{Kim80,Ban86,Iva90,Dec93,Fin94}:
\be
   {\cal M}[\tau^-(s) \to \pi^-(p) \nu_\tau(q) \gamma(k)] =
   {\cal M}_{IB} + {\cal M}_{SD}
\ee
The internal bremsstrahlung corresponds to hooking photons to the external
lines, considering the pion as an structureless elementary particle:
\begin{eqnarray}
{
   {\cal M}_{IB}  =  - G_F \cos \theta_c e f_\pi m_\tau
   \bar{u}_\nu(q) \gamma_+}
   \left[ \frac{p \cdot \epsilon}{p \cdot k}
   + \frac{\strich{k}\strich{\epsilon}}{2 s \cdot k}
   - \frac{s \cdot \epsilon}{s \cdot k} \right]
   u_\tau(s)
\end{eqnarray}
The structure dependent radiation can be defined as the contribution from
the diagram in Fig.~\ref{fig2}(e) minus
the effective point pion contribution, resulting in
\begin{eqnarray}
{   \mcal_{SD}  =   - \frac{G_F \cos \theta_c}{\sqrt{2}} }
     \Bigg\{ i \epsilon_{\mu \nu \rho
      \sigma}
      \big[ \bar{u}_\nu \gamma^\mu \gamma_- u_\tau \big]
     \epsilon^\nu k^\rho p^\sigma \frac{F_V(t)}{m_\pi}
\nonumber \\
   \qquad \qquad \qquad
+ \bar{u}_\nu(q) \gamma_+ \left[ (p \cdot k) \strich{\epsilon} -
      (\epsilon \cdot p) \strich{k} \right] u_\tau(s)
     \frac{F_A(t)}{m_\pi}
      \Bigg\}
\end{eqnarray}
This amplitude ${\cal M}_{SD}$ involves two form
factors $F_V(t)$ and $F_A(t)$, which depend on the invariant mass
squared of the pion-photon system, $t = (p + k)^2$.
The vector and axial vector form factors $F_V$ and $F_A$ correspond to
the $J^P = 1^-$ and $J^P = 1^+$ projections of the $W$ boson, respectively.
And so in order to be able to calculate the decay rate,
we have to parameterize the form factors $F_V(t)$ and $F_A(t)$.
%
\subsection{Normalization at $t=0$}
%
We use the following values for the form factors at zero momentum transfer:
\begin{eqnarray}
   F_V(t=0) & = \frac{\displaystyle m_\pi}{\displaystyle 4 \sqrt{2} \pi^2
      f_\pi}
   = 0.0270
\nonumber \\
   F_A(t=0) & = 0.0116 \pm 0.0016
\end{eqnarray}
Here $F_V(0)$ is obtained from the axial anomaly, whereas $F_A(0)$ has
been measured in the radiative pion decay $\pi\to e \nu_e \gamma$
\cite{RPP92}.

However, for the radiative tau decay, we need $F_V$ and $F_A$ in the
whole
range $m_\pi^2 \leq t \leq m_\tau^2$. In order to parameterize the energy
dependence of the form factors we assume vector and axial vector meson
dominance.
%
\subsection{Parametrization of $F_A(t)$}
%
In the case of the axial form factor, the only established resonance which
has the correct quantum numbers is the $a_1(1260)$. Thus we assume
\be
   F_A(t) = \mbox{BW}_{a_1}(t) F_A(0)
\ee
where in the normalized Breit Wigner resonance factor
\be
   \mbox{BW}_{a_1}(t) = \frac{m_{a_1}^2}
   {m_{a_1}^2 - t - i m_{a_1} \Gamma_{a_1}(t)}
\ee
we use an energy dependent width $\Gamma_{a_1}(t)$ as calculated from
the $a_1 \to 3 \pi$ and $a_1 \to \rho \pi$ phase space \cite{Kue90}.
\subsection{Parametrization of $F_V$}
In the case of the vector form factor, there are three resonances with
the correct quantum numbers in the mass range covered by the tau, viz.\
the $\rho = \rho(770)$, $\rho' = \rho(1450)$ and the $\rho'' =
\rho(1700)$. And so the vector meson dominance ansatz for $F_V(t)$ reads
\begin{eqnarray}
{F_V(t) = \frac{F_V(0)}{1 + \lambda + \mu}}
   \Big[ \mbox{BW}_\rho(t) + \lambda \mbox{BW}_{\rho'}(t)
   + \mu \mbox{BW}_{\rho''}(t) \Big]
\end{eqnarray}
We have to fix the two parameters $\mu$ and $\lambda$ which determined
the size of the contribution of the higher radial resonances.
Noting that $F_V(t)$ is related by CVC (conserved vector current) to the
form factor $F_{\pi\gamma\gamma^\star}(t)$ which determines the
coupling $\pi^0 \gamma \gamma^\star(q^2 = t)$ \cite{Vak58},
we are able to find four
constraints for the two parameters:
\begin{enumerate}
\item The limit of $F_{\pi\gamma\gamma^\star}$ for large negative $t$
is predicted by perturbative QCD \cite{Lep79}
\be
   \lim_{t \to - \infty} t F_{\pi\gamma\gamma^\star}(t) =  2 f_\pi
\ee
\item The slope of $F_{\pi\gamma\gamma^\star} (t)$ at $t=0$ has been
measured in $\pi^0 \to e^+ e^- \gamma$ \cite{RPP92}.
\item
The value $F_V(m_\rho^2)$ of the vector form factor at the rho mass is
related to the decay width $\Gamma(\rho \to \pi \gamma)$ which has been
measured.
\item
Similarly, $F_V(m_{\rho'}^2)$ is related to $\Gamma(\rho' \to \pi
\gamma)$, which has not been measured but is in turn related by vector
meson dominance to the
measured width $\Gamma(\rho' \to \pi \omega)$.
\end{enumerate}
The parameter choice
\begin{eqnarray}
   \lambda = 0.136; \qquad   \mu     = - 0.051
\end{eqnarray}
simultaneously fulfills all four constraints.
%
\subsection{Numerical Results}
%
\begin{figure}
\caption{Photon energy spectrum of the decay $\tau\to\pi\nu_\tau\gamma$:
Total prediction (solid), pure internal bremsstrahlung
(dashed) and the prediction from TAUOLA + PHOTOS
(dots with statistical error bars)}
\label{figu4}
\begin{center}
\fbox{   \epsfig{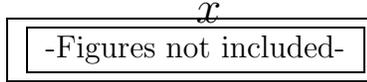}}
\unitlength1.0cm
\begin{picture}(0,0)
    \put(-10.,18.25){\makebox(0,0)[l]{{\bf (a)}
    Spectrum for $0.1 \leq x \leq 1$}}
    \put(-10.,9.2){\makebox(0,0)[l]{{\bf (b)} Close-up of (a) with $0.5%
\leq x \leq 1$ }}
   \put(-2.5,9.7){\makebox(0,0)[l]{\Large  $\D x $ }}
   \put(-5.,16.){\makebox(0,0)[l]{\Large $ \D\frac{ 10^{3}}{ \gnr}%
   \frac{d\Gamma}{dx}$}}
   \put(-2.5,0.6){\makebox(0,0)[l]{\Large  $\D x $ }}
    \put(-5,6.5){\makebox(0,0)[l]{\Large $ \D\frac{ 10^{3}}{ \gnr}%
    \frac{d\Gamma}{dx}$}}
%
\end{picture}
  \end{center}
\end{figure}
Having obtained parameterizations for the hadronic structure dependent
form factors, we can calculate decay distributions for
$\tau\to\pi\nu_\tau\gamma$ \cite{Dec93}.
In Fig.~\ref{figu4} we display the photon
energy spectrum $d \Gamma / dx$ of the decay.
The dimensionless
parameter $x$ is defined by
\be
   x = \frac{2 E_\gamma}{m_\tau}
\ee
where $E_\gamma$ is the photon energy in the tau rest frame.
In Fig.~\ref{figu4}, we do not only display the total spectrum, which we
predict, but also the contribution from pure internal bremsstrahlung (IB)
(i.e.\ neglecting hadronic structure dependent effects) and the result
from the Monte-Carlo program PHOTOS \cite{Bar90}.
PHOTOS provides a semiclassical
approximation of the IB contribution. We find from
Fig.~\ref{figu4} that PHOTOS gives a very good approximation to the IB
part. Furthermore we find that the IB is strongly dominating the total
spectrum, except for the region of very hard photons where the rate is
extremely small anyway.

So in fact we find that PHOTOS gives a very good approximation to
the photon spectrum of $\tau\to\pi\nu_\tau\gamma$,
and that hadronic structure dependent effects are almost invisible in
this spectrum.


A observable much better suited to separate the structure dependent effects
from the internal bremsstrahlung is
the pion-photon invariant mass spectrum $d\Gamma/dm_{\pi\gamma}$, which
we display in Fig.~\ref{figu5}.
The dimensionless parameter $z$ is defined by
\be
    z = \frac{m_{\pi\gamma}^2}{m_\tau^2}
\ee
In this spectrum, a clear $\rho$ resonance
peak and a smaller $a_1$ peak are visible.

A measurement of this spectrum would be very interesting both in order
to check our model and in order to have another channel
to measure the poorly known width $\Gamma_{a_1}$.
However, there is a
very large background from the decay
$\tau\to\pi^-\pi^0\nu_\tau$, where the $\pi^0$ decays into two photons
and one of these may escape detection.

It is interesting to mention the total size of the hadronic structure
dependent contribution (integrated over the full phase space),
\begin{eqnarray}
{  \frac{\Gamma_{SD + INT}(\tau\to\pi\nu_\tau\gamma)}
  {\Gamma_0(\tau\to\pi\nu_\tau)}
   = (0.11 - 0.06) \% }
  = 0.05 \%
\end{eqnarray}
where the first number (0.11) corresponds to the structure dependent
amplitude squared (SD) and the second one (-0.06) to the interference between
structure dependent radiation and internal bremsstrahlung (INT).

\begin{figure}
\caption{Pion-photon invariant mass spectrum of the
   decay $\tau\to\pi\nu_\tau\gamma$: Internal bremsstrahlung (IB,
   dashed),
  structure dependent radiation (SD, dash-dotted),
  interference of IB and SD (INT, dotted),
  and the total spectrum (solid)}
\label{figb6}
\begin{center}
\fbox{   \epsfig{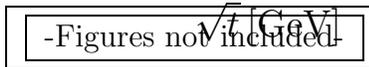}}
\unitlength1.0cm
\begin{picture}(0,0)
   \put(-13,18.1){\makebox(0,0)[l]{{\bf (a)}
   $0.4 \,\mbox{GeV} \leq \sqrt{t} \leq m_\tau$
}}
   \put(-12.5,9.2){\makebox(0,0)[l]{{\bf (b)} Close-up of (a) %
   with $\D 1.0\, \mbox{GeV} \leq \sqrt{t} \leq m_\tau$}}
  \put(-2.5,9.6){\makebox(0,0)[l]{\large  $\D \sqrt{t}\, \mbox{[GeV]} $ }}
  \put(-5,16.){\makebox(0,0)[l]{\Large $ \D\frac{ 10^{3}}{ \gnr}%
  \frac{d\Gamma}{dz}$}}
  \put(-2.5,0.3){\makebox(0,0)[l]{\large  $\D \sqrt{t}\, \mbox{[GeV]} $ }}
  \put(-5,7.){\makebox(0,0)[l]{\Large $ \D\frac{ 10^{3}}{ \gnr}%
  \frac{d\Gamma}{dz}$}}
%
\end{picture}
  \end{center}
\end{figure}
%
\section{Virtual Corrections}
%
%
Calculating the virtual corrections, we have to perform loop
integrations over the Euclidean photon momentum $k_E^2$ from $k_E^2
= 0$ up to $k_E^2 = m_Z^2$. In order to be able to perform this
integration, we separate the integration range into two parts
\cite{Sir72,Hol90,Dec94}.
In the long distance region of $k_E^2 = 0 \cdots
\mu_{cut}^2 \approx O(1 \, \mbox{GeV}^2)$, mesons are the relevant degrees of
freedom. For this region we build a realistic
phenomenological model, taking into account low energy theorems of QCD,
assuming vector meson dominance and considering experimental constraints
such as the electromagnetic form factor of the pion, which is known
quite well in the relevant momentum region.
For the short distance integration, $k_E^2 = \mu_{cut}^2 \cdots m_Z^2$,
we use the parton model.
\subsection{Long Distance Part}
\begin{figure}
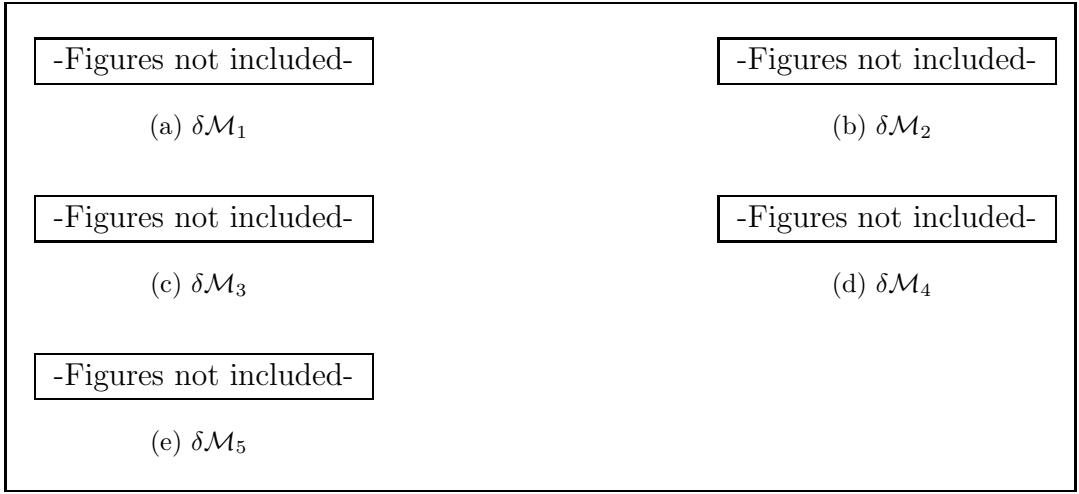

   \caption{Effective Point pion diagrams}
   \label{figc4}
\begin{center}
\fbox{\parbox{14cm}{
\subfigure[$ \delta \mcal_1$]{
\mbox{   \epsfig{figure=figc4a.ps,width=4.6cm,%
bbllx=2cm,bblly=7cm,bburx=18.1cm,bbury=21.6cm}}
}
\subfigure[$\delta \mcal_2$]{
\mbox{   \epsfig{figure=figc4b.ps,width=4.6cm,%
bbllx=2cm,bblly=7cm,bburx=18.1cm,bbury=21.6cm}}
}
\subfigure[$\delta \mcal_3$]{
\mbox{   \epsfig{figure=figc4c.ps,width=4.6cm,%
bbllx=2cm,bblly=7cm,bburx=18.1cm,bbury=21.6cm}}
}
\subfigure[$\delta \mcal_4$]{
\mbox{   \epsfig{figure=figc4d.ps,width=4.6cm,%
bbllx=2cm,bblly=7cm,bburx=18.1cm,bbury=21.6cm}}
}
\subfigure[$ \delta \mcal_5$]{
\mbox{   \epsfig{figure=figc4e.ps,width=4.6cm,%
bbllx=2cm,bblly=7cm,bburx=18.1cm,bbury=21.6cm}}
}
%
}}
  \end{center}
\end{figure}

\begin{figure}
\caption{Vector meson dominance of coupling of the photon to the pion.
In fact we include the $\rho$, the $\rho'$ and the $\rho''$ with
relative strengths which have been obtained by fitting the measured
pion electromagnetic form factor \protect\cite{Kue90}.}
\label{figc8}
\begin{center}
\fbox{\epsfig{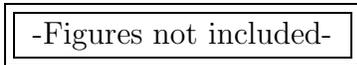}}
\end{center}
\end{figure}

\begin{figure}
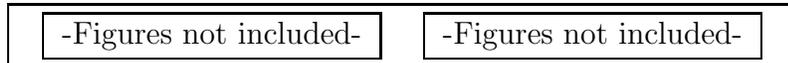

   \caption{
   Hadronic structure dependent loop diagrams}
   \label{figc9}
\begin{center}
\fbox{\begin{tabular}[t]{cc}
%
\mbox{   \epsfig{figure=figc9a.ps,width=5cm,%
bbllx=1.9cm,bblly=8.cm,bburx=18.3cm,bbury=21cm}}
&
\mbox{   \epsfig{figure=figc9b.ps,width=5cm,%
bbllx=1.9cm,bblly=8.cm,bburx=18.3cm,bbury=21cm}}
\end{tabular}
}
  \end{center}
\end{figure}

In the integration over small photon momenta, $k_E^2 = 0 \cdots
\mu_{cut}^2$, we proceed as follows. We start from the diagrams which
correspond to the lowest order, $O(P^2)$, of low energy QCD (effective
point pion \cite{Dec93b},
see Fig.~\ref{figc4}). We then modify these diagrams to
account for the momentum dependence of the electromagnetic form factor
of the pion, which is known to be dominated by rho-like resonances,
see Fig.~\ref{figc8}.
We then add to this the loop diagrams which correspond to
the structure dependent (SD) part in the radiative decays, see
Fig.~\ref{figc9}. As indicated in the figure, we assume a double vector
meson dominance of the relevant form factors $H_V$ and $H_A$,
\begin{eqnarray}
\label{eqndouble}
  H_V(k,p) & = \mbox{BW}_\omega(k^2) F_V[(k+p)^2]
\nonumber \\
  H_A(k,p) & = \mbox{BW}_\rho(k^2) F_A[(k+p)^2]
\end{eqnarray}
where $k$ is the momentum of the virtual photon, $p$ the pion
momentum, and $F_V$ and $F_A$ are the form factors involved in the
radiative decays (real photons).
In order to estimate the model dependence, we also
compare to a form with a single vector meson dominance,
\begin{eqnarray}
\label{eqnsingle}
  H_V(k,p) & = F_V[(k+p)^2]
\nonumber \\
  H_A(k,p) & = F_A[(k+p)^2]
\end{eqnarray}
%
%
\subsection{Short Distance Corrections}
\begin{figure}
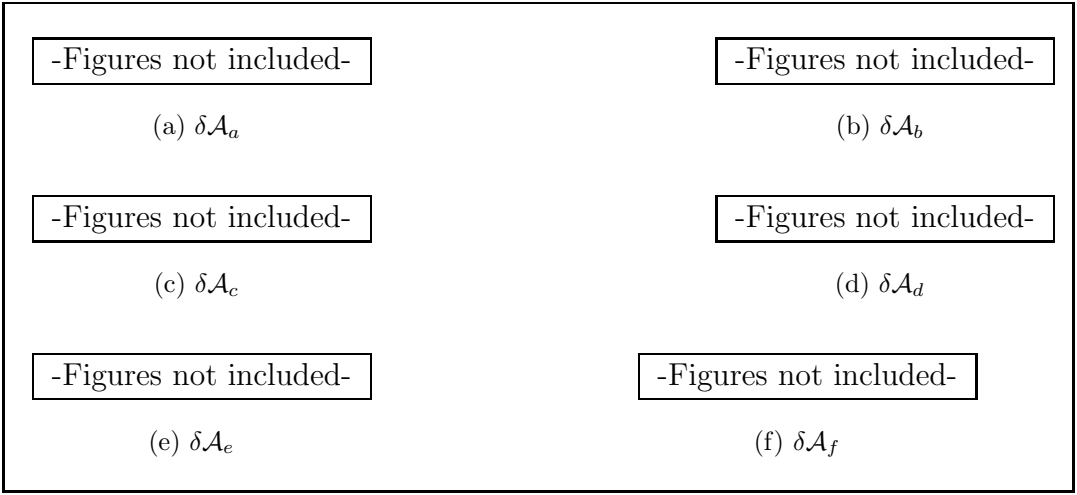

   \caption{Short distance diagrams}
   \label{figc6}
\begin{center}
\fbox{\parbox{14cm}{
\subfigure[$ \delta {\cal A}_a$]{
\mbox{   \epsfig{figure=figc6a.ps,width=4.6cm,%
bbllx=3.5cm,bblly=8.7cm,bburx=17.8cm,bbury=21.5cm}}
}
\subfigure[$\delta {\cal A}_b$]{
\mbox{   \epsfig{figure=figc6b.ps,width=4.6cm,%
bbllx=3.5cm,bblly=8.7cm,bburx=17.8cm,bbury=21.5cm}}
}
\subfigure[$\delta {\cal A}_c$]{
\mbox{   \epsfig{figure=figc6c.ps,width=4.6cm,%
bbllx=3.5cm,bblly=8.7cm,bburx=17.8cm,bbury=21.5cm}}
}
\subfigure[$\delta {\cal A}_d$]{
\mbox{   \epsfig{figure=figc6d.ps,width=4.6cm,%
bbllx=3.5cm,bblly=8.7cm,bburx=17.8cm,bbury=21.5cm}}
}
\subfigure[$ \delta {\cal A}_e$]{
\mbox{   \epsfig{figure=figc6e.ps,width=4.6cm,%
bbllx=3.5cm,bblly=8.7cm,bburx=17.8cm,bbury=21.5cm}}}
\mbox{\hspace*{3cm}}
\subfigure[$ \delta {\cal A}_f$]{
\mbox{   \epsfig{figure=figc6f.ps,width=4.6cm,%
bbllx=3.5cm,bblly=8.7cm,bburx=17.8cm,bbury=21.5cm}}}}}
  \end{center}
\end{figure}

%
For large virtual photon momenta, $k_E^2 = \mu_{cut}^2 \cdots m_Z^2$, we
calculate the corrections to the elementary vertex $\tau \to \nu_\tau
\bar{u} d$, see Fig.~\ref{figc6}, and plug the result into
Fig.~\ref{figw}. Doing the same thing for the pion decay, we obtain the
short distance contribution to the radiative correction to the ratio
$R_{\tau/\pi}$ in the form
\be
  \Big( \delta R_{\tau/\pi} \Big)_{sd} = \frac{3}{2 f_\pi}
   \int_{-1}^{+1} du \, \Phi_\pi(u) r(u)
\ee
The integration extends over the scaled relative momentum of the two quarks in
the infinite momentum frame. $\Phi_\pi(u)$ is an unknown parton
distribution function, and $r(u)$ has been calculated from the short
distance diagrams. $r(u)$ varies only very little over the integration
range, and so we can approximate it by its value at $u=0$,
\be
  \Big( \delta R_{\tau/\pi} \Big)_{sd} \approx r(0) \frac{3}{2 f_\pi}
   \int_{-1}^{+1} du \, \Phi_\pi(u) = r(0)
\ee
where last equation follows from a sum rule \cite{Bra70}.
In fact we find
that the short distance correction is dominated by a leading logarithmic
contribution \cite{Dec94,Sir82},
\be
  \Big( \delta R_{\tau/\pi} \Big)_{sd} \approx
  \frac{2 \alpha}{\pi} \frac{m_\tau^2}{m_\tau^2 - \mu_{cut}^2} \ln
  \frac{\mu_{cut}}{m_\tau}
\ee
\begin{figure}
\caption{Decay $\tau\to\pi\nu_\tau$ via an intermediate
quark-antiquark state.  The bubble on the left hand side is to be
replaced by the short
distance diagrams $\delta {\cal A}_i$ of Fig.~\protect\ref{figc6}}
\label{figw}
\begin{center}
\fbox{\epsfig{figure=figwave.ps,width=7cm,%
bbllx=1cm,bblly=10cm,bburx=20.5cm,bbury=19.8cm}}
\end{center}
\end{figure}
%
\subsection{Numerical Results}
%
Adding up long and short distance corrections, we obtain the total
radiative correction, which depends on the choice of the matching scale
$\mu_{cut}$ and on the choice of the parameters of the hadronics. In
Fig.~\ref{figu6} we display the radiative correction to the width
$\tau\to\pi\nu_\tau$ in variation with $\mu_{cut}$, using three
different sets for the hadronic parameters. There is a standard set (I)
with some central values, and sets (II) and (III) with reasonable
variations around these central values. While for the sets (I) and (III)
the dependence of the correction of the matching scale $\mu_{cut}$ is
reasonably small, it is unacceptably large for (II). In fact ,(I) and
(III) use the double vector meson dominance parameterization of
(\ref{eqndouble}), (II), however,
uses the single pole form of (\ref{eqnsingle}),
which therefore can be excluded.

We obtain similar results for the radiative correction to $\Gamma(
\pi\to\mu\nu_\mu)$. Taking the ratio, we find
\begin{eqnarray}
{  \delta R_{\tau/\pi} = (0.16 \pm 0.08 \pm 0.04 \pm 0.07) \%}
   = (0.16 \pm 0.14 ) \%
\end{eqnarray}
The first error quoted (0.08\%) arises from the matching uncertainties
in $\Gamma(\tau\to\pi\nu_\tau(\gamma)$, the second one from the matching
in $\Gamma(\pi\to\mu\nu_\mu(\gamma)$, and the last one from the
uncertainties in the hadronic parameters.
The number given here for $\delta R_{\tau/\pi}$,  $(0.16 \pm 0.14) \%$,
differs slightly from the number we quoted in \cite{Dec94}, $\delta
R_{\tau/\pi} = \left( 0.16_{-0.14}^{+0.09} \right)\%$.
This is due to the fact that in
\cite{Dec94}, we estimated the matching uncertainty by considering the
variation of the ratio $R_{\tau/\pi}$, whereas now we have discussed the
matching uncertainties of the individual decay rates. The latter
approach should give a better estimate of the true model dependence,
because in the ratio scale dependences associated with a mismatch of
long and short distances tend to cancel at large scales.

Note that the rather small
total radiative correction arises from cancellations of larger numbers
with opposite signs.

Similarly, we obtain the radiative correction to the ratio of the tau
decay into a kaon $\Gamma(\tau\to K \nu_\tau(\gamma))$ and the decay
width of the kaon $\Gamma(K \to \mu\nu_\mu (\gamma)$
\begin{eqnarray}
{
  \delta R_{\tau/K} = (0.90 \pm 0.08 \pm 0.09 \pm 0.14) \%}
   = (0.90 \pm 0.22 ) \%
\end{eqnarray}

Our result for $R_{\tau/\pi}$ should be compared with the recent
estimate by Marciano and Sirlin \cite{Mar93}, which in terms of $R_{\tau/\pi}$
reads
\be
   \delta R_{\tau/\pi} = (0.67 \pm 1.)\%
\ee
where the $\pm 1.\%$ is the authors' estimate of the missing long
distance corrections to the tau decay rate, which they did not
calculate. And so we confirm their result within their estimated error
bars, but we are able to reduce the error substantially by performing a
complete calculation.
\begin{figure}
   \label{figu6}
   \caption{Radiative correction to $\Gamma(\tau\to\pi\nu_\tau)$, using
different choices for the parameters of the hadronic structure dependent
correction: Standard choice (I) (solid), and variations (II) and (III)
(dashed and dotted, respectively)}
\bildchen{zeigtauo.ps}%
{\frac{\D\delta \Gamma}{\D \Gamma_0} (\tau\to\pi \nu_\tau(\gamma))[\%]}%
{ \mu_{cut}\, [\mbox{GeV}]}
\end{figure}
Translating the radiative corrections into predictions for the
branching ratios, we get
\begin{eqnarray}
{  \mbox{BR}(\tau\to\pi\nu_\tau(\gamma))  =  }
   (10.946 +
   \overbrace{0.005}^{\mbox{\tiny SD + INT}} \pm 0.020) \%
   \times \left(\frac{\tau_\tau}{291.6 \, \mbox{fs}}\right)
\nonumber \\
\nonumber \\
{   \mbox{BR}(\tau\to K \nu_\tau(\gamma))  =  }
   (0.723 +
   \overbrace{0.002}^{\mbox{\tiny SD + INT}} \pm 0.004) \%
   \times \left(\frac{\tau_\tau}{291.6 \, \mbox{fs}}\right)
\nonumber \\
\nonumber \\
{   \mbox{BR}(\tau\to h \nu_\tau(\gamma))    =  }
   (11.669 +
   \overbrace{0.007}^{\mbox{\tiny SD + INT}} \pm 0.021) \%
   \times \left(\frac{\tau_\tau}{291.6 \, \mbox{fs}}\right)
\end{eqnarray}
where $h$ denotes the inclusive sum of pions and kaons.
Note that here we have separated out the contribution
SD + INT associated with
structure dependent radiation, corresponding to the decay chains
$\tau\to\rho\nu_\tau$, $\rho\to\pi\gamma$ and $\tau\to a_1 \nu_\tau$,
$a_1 \to \pi\gamma$. Whereas we have included all photons
in $\delta R_{\tau/\pi}$, experimental numbers for the branching ratios
do not include hard photons.
Note, however, that the size of this
SD + INT part is in all three cases
small compared to the overall uncertainty of the prediction.
Using the new world average for the tau
lifetime \cite{taulifetime}
\be
   \tau_\tau = (291.6 \pm 1.6 ) \, \mbox{fs}
\ee
we obtain
\begin{eqnarray}
   \mbox{BR}_\pi^{theo} & = & (10.95 \pm 0.06) \%
\nonumber \\
   \mbox{BR}_K^{theo} & = & (0.723 \pm 0.006) \%
\nonumber \\
   \mbox{BR}_h^{theo}   & = & (11.67 \pm 0.06) \%
\end{eqnarray}
These predictions agree within one standard deviation with the new
world averages \cite{taubrs} quoted at this conference.
%
\section*{Acknowledgement}
M.F. would like to acknowledge helpful and illuminating discussions with
J.\ Bijnens,
J.H.\ K\"uhn and with A.\ Sirlin, which have been important
for the work reported here.
He would also like to acknowledge financial support by the
Graduiertenkolleg Elementarteilchenphysik Karlsruhe and by the HCM
program under EEC contract number CHRX-CT920026.

\end{document}